\documentclass{eas}
\usepackage{graphicx}
\usepackage{natbib}
\begin{document}

\title{Chemical and Dynamical Properties of the Stellar Halo} 
\author{Chris Brook}\address{Dept. of Astronomy, University of Washington, Box 351580, Seattle, WA 98195, USA}\secondaddress{D\'{e}partement de physique, de g\'enie physique et d'optique, Universit\'{e} Laval, Qu\'{e}bec, Qc, G1K 7P4, CANADA}
\author{Daisuke Kawata}\address{The Observatories of the Carnegie Institution of Washington, 
813 Santa Barbara St., Pasadena, CA 91101}
\author{Hugo Martel}\sameaddress{2}
\author{Brad Gibson}\address{Centre for Astrophysics, University of Central Lancashire, Preston, PR1, 2HE, UK}
\author{Evan Scannapieco}\address{Kavli Institute for Theoretical Physics, Kohn Hall, UC Santa Barbara, Santa Barbara, CA 93106}
\begin{abstract}
The difference in density profiles of the contributions from different density peaks to dark matter halos results in certain expectations about the Milky Way's stellar halo. We cut our simulated halo stars into two populations: those forming before/during the last major merger, and those accreted after  the last major merger.  The former population are  more centrally located at $z=0$, while stars forming in low mass late forming proto-galaxies are spread through the halo. A difference in observed  binding energy  distinguishes these two populations.  We look at possible chemical abundance signatures of the two populations. We also show that galaxies forming in isolated low $\sigma$ peaks will form from primordial material. Thus, even though the oldest stars are centrally concentrated as they originated in the early collapsing, densest regions, primordial stars would be found distributed throughout the halo. Thus, the lack of observed metal free stars can be taken as directly constraining the Population III IMF, and the lowest metallicity observed stars  can be interpreted as holding clues to the chemical yields of Pop III stars.

\end{abstract}
\maketitle
\section{Introduction}
As it contains very old stars, some almost as old as the universe itself, the stellar halo of our own Galaxy holds valuable clues to galaxy formation.  Deconstructing the formation and evolution
of the Milky Way in a star-by-star fashion is a key driver behind the 
development facilities of such as the European
Space Agency's GAIA mission, NASA's Space Interferometry Mission,
and ongoing projects, such as 
Sloan Extension for Galactic Understanding and Exploration
(SEGUE) and the RAdial Velocity Experiment (RAVE).
A seminal view of the field of {\it{Galactic Archaeology}} is
found in \cite{fb02}. 

Here we look at some of the signatures of the hierarchical clustering of ``building blocks'', which Cold Dark Matter cosmological theories predict played a  central role in the assembly history of the Milky Way. 
We present our results as being intimately linked to the build up of dark matter halos, in particular, the flatter density profile from  the contribution of matter from low $\sigma$ density peaks compared to high $\sigma$ peaks.  We do so with the aid of GCD+, a chemo-dynamical galaxy evolution code.

\section{Low Mass Stellar Halo in Simulations}
\begin{figure}
\includegraphics[width=7.in]{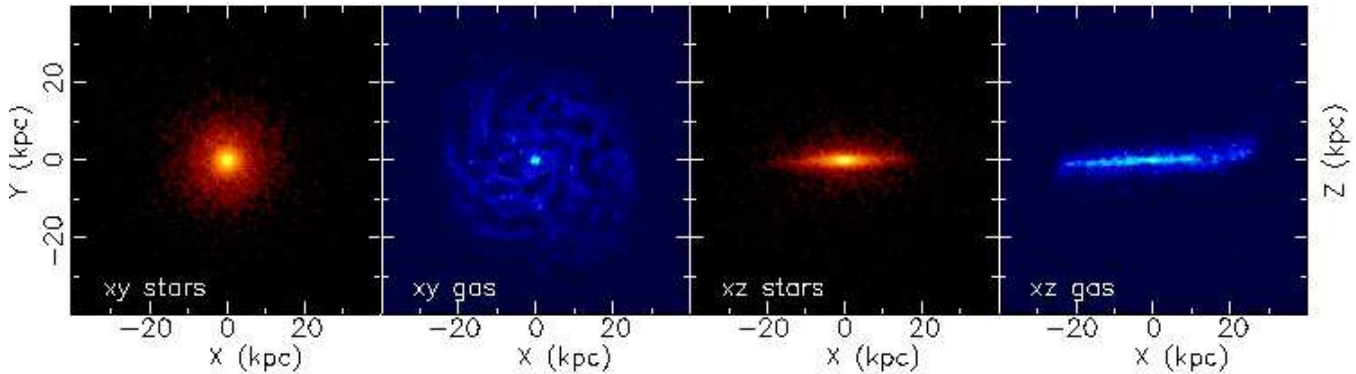}
\caption{A density map of our simulated galaxy is shown face-on (left panels) and edge-on, for stars and gas.  }  \label{final}
\end{figure}

GCD+ self-consistently follows the effects of gravity, gas dynamics, radiative cooling, and star formation. GCD+ also takes into account metal enrichment and energy released by both Type II and Type Ia  supernovae, as well as the metal enrichment from intermediate mass stars. Full details of GCD+ can be found in \cite{kawata}. We apply the Adiabatic Feedback Model described in \cite{brook04}. The model assumes that the gas within the SPH smoothing kernel of SNe II explosions has an adiabatic phase.
Energy feedback results in  regulation of  star formation in low mass proto-galaxies,  crucial in forming disk dominated galaxies \citep{brook04}.  This regulation of star formation is also required to ensure that the disk retains enough angular momentum to match observations \citep{thacker,governato}, and  in N-body/semi-analytic models of stellar halo formation in order to explain metallicities and abundances \citep{moore,bullock}.  The stars and gas of  our final galaxy are shown in figure.~\ref{final}, both face-on (left panels) and edge-on (right panels).

\section{In-situ versus accretion: motivation and signatures}
\begin{figure}
\includegraphics[width=6.5in]{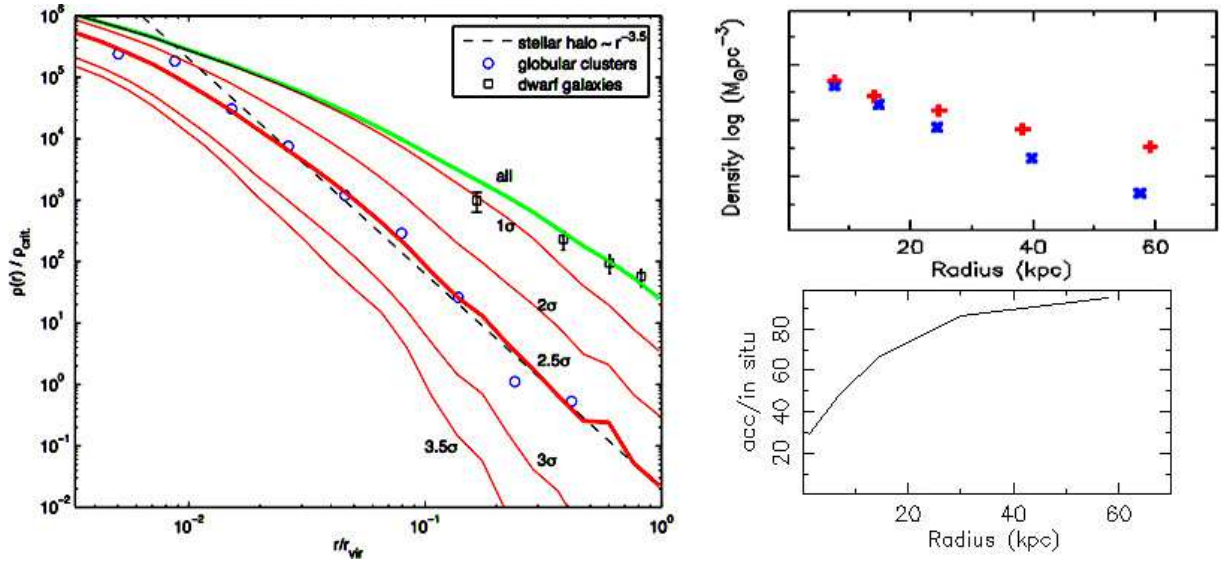}
\caption{Left panel from \cite{moore}: the radial distribution of old stellar systems compared with rare peaks at  z = 0 in LCDM galaxy. The thick
green curve is the total mass distribution today,  red
curves show the present day distribution of material that collapsed into 1, 2, 2.5, 3 and 3.5$\sigma$  peaks at a redshift $z = 12$, 
circles show the observed spatial distribution of the Milky WayÕs
old metal poor globular cluster system. The dashed line indicates
a power law  $\rho(r) \propto r^{-3.5}$ which represents the old halo stellar
population. The squares show the radial distribution of surviving 2.5$\sigma$ peaks which are slightly more extended than the overall
NFW like mass distribution, in good agreement with the observed
spatial distribution of the Milky WayÕs satellites. The top right panel shows the contribution to the  density profile of stellar halo stars in our simulation, coming from stars formed before and during the last major merger (``in-situ" stars, blue) and stars accreted after this event (``accreted" stars, red). The lower right panel shows the ratio of accreted to in-situ halo stars as a function of radius. }   \label{moore}
\end{figure}

The CDM density distribution has substructure on different  scales. These contribute in different ways to the final dark matter halo, with matter originating  in high $\sigma$ peaks being centrally concentrated compared with matter originating in  low $\sigma$ peaks. This is shown clearly in the left panel of figure~\ref{moore},  taken from figure~2 of \cite{moore}, and which shows the contribution  to a Milky Way sized dark halo from 1,2, \& 3$\sigma$ over-densities at z=12.

In order to examine possible  effects of this on the stellar halo of the Milky Way, we have divided the stellar halo of our simulated Milky Way in to two populations: {\it{in-situ}} and {\it{accreted}}. We define these populations as; stars within the central  galaxy at the end of  the last major merger (in-situ); and those accreted after the last major merger (accreted). The top right panel of figure~\ref{moore} shows the contribution to the final stellar halo density profile of these two populations. Clearly, in-situ stars (blue ``$\times$") have a steeper density profile than accreted stars (red ``$+$"). In the lower panel of figure~\ref{moore}, we plot the relative contribution of the accreted to in-situ populations as a function of radius. In the bulge, the in-situ population dominates; at the solar neighborhood, the two populations contribute similar numbers of stars; in outer regions, the accreted stars dominate. The details of this plot, of course, are dependent on the unique merging history of this particular simulated Milk Way. It is notable, however, the similarities to figure~2  of \cite{abadi}, even though their model differs to ours in merging history, feedback implementation, and stellar halo mass.

\begin{figure}
\includegraphics[width=7.8in]{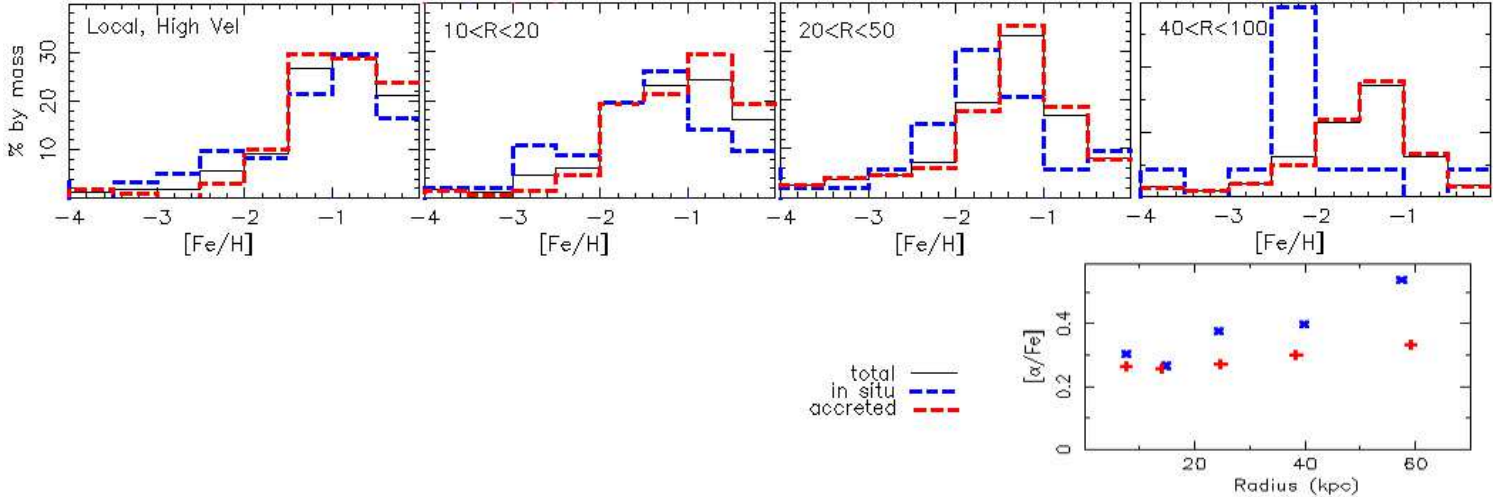}
\caption{Metallicity distribution functions (MDF) of the accreted (red) and in-situ (blue) populations.  
The left panel plots ``solar neighborhood" halo stars, defined as the region $5<R_{XY}<11$ and $|Z|<1$, and halo stars satisfy the condition $\sqrt{T^2+(V-V_{ave})^2}>V_{ave}^2$, where  $V$, $U$, $W$ are tangential, radial, and out of the plane velocities respectively, $T^2=U^2+W^2$ and $V_{ave}$ is the rotation velocity of the disk. The next three panels select halo stars by region, taking stars with $|Z|>4$, and  regions as indicated. Difference in the peak of the MDFs begin to appear at larger radii.  The lower right panel  shows the [$\alpha$/Fe] abundances, for the two populations at different radii. }  \label{metals}
\end{figure}

It is interesting to then ask whether these two populations will differ in their metallicities. And whether this has implications for comparing stellar halo stars selected in different ways by observers, kinematically selected local stars, compared with spatially selected.  In figure~\ref{metals} we plot the metallicity distribution functions (MDF) of the two populations using different methods to select stellar halo stars, with in-situ (accreted) stars having blue (red) lines. 
The left panel selects ``solar neighborhood" stars kinematically. The solar neighborhood is defined as the region $5<R_{XY}<11$ and $|Z|<1$, and stars are selected from the Toomre diagram, as satisfying the condition $\sqrt{T^2+(V-V_{ave})^2}>V_{ave}^2$, where  $V$, $U$, $W$ are tangential, radial, and out of the plane velocities respectively, $T^2=U^2+W^2$ and $V_{ave}$ is the rotation velocity of the disk. The next three panels select halo stars by region, taking stars with $|Z|>4$, and then the regions where $10<R_{XYZ}<20$, $20<R_{XYZ}<50$, $40<R_{XYZ}<100$. Difference in the peak of the MDFs begin to appear at larger radii. The in-situ population show a greater metallicity gradient than the accreted population, with lower metallicity stars expected at higher radii. The lower panel of figure~\ref{metals} shows the [$\alpha$/Fe] abundances, for the two populations at different radii. Again, the difference between the two populations becomes greater at large radius, and in-situ stars have a greater radial gradient. The abundances are very similar in the solar region.

\begin{figure}
\includegraphics[width=6.in]{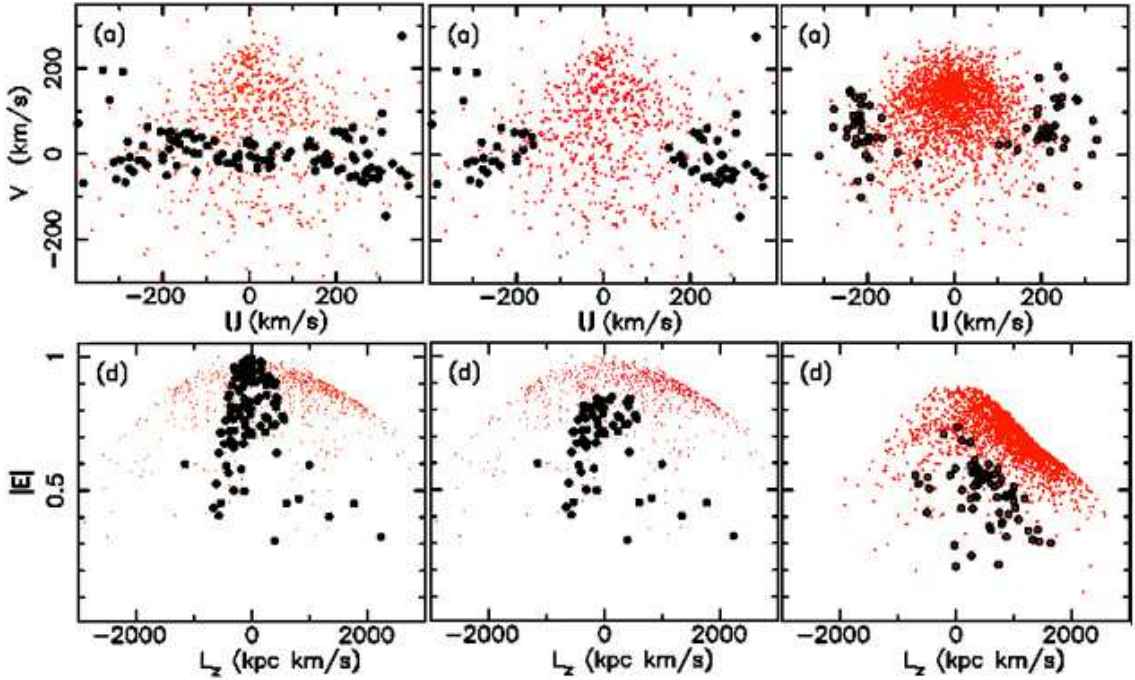}
\caption{Left panel shows   observed solar neighborhood halo stars from the catalogue of \cite{cb00} in the U-V  (top) and total energy (E) - angular momentum (L$_Z$) planes. We highlight stars with   $e>0.8$ and [Fe/H]$\sim -1.7$.   A cut in $|E|$ is made in the observed data, with low bound stars shown in the middle panel.   Velocity space plots of these low bound observed  stars show similar kinematic properties to the accreted stars in the simulation, shown in the right hand panel. }  \label{phase}
\end{figure}

Accreted  halo stars can also be expected to be less bound than in-situ stars.  In the left panel of figure~\ref{phase}, we plot phase space of stars from the catalogue of \cite{cb00} of observed solar neighborhood halo stars in the U-V plane (top) and total energy (E)  versus angular momentum (L$_Z$) planes (red). We highlight (black circles) stars which belong to a dense clump in eccentricity($e$)-metallicity plane, corresponding to $e>0.8$ and [Fe/H]$\sim -1.7$.  The right hand panel shows accreted simulation stars.  Note that they have low $|E|$ (are less bound). This motivates us to make a cut in $|E|$ for the observed data, in order to separate accreted stars from in-situ ones; low bound stars are shown in the middle panel.  The corresponding velocity space plots of these low bound stars show similarities to the accreted stars in the simulation. 
This is true for the U-W and U-V plots as well, not shown here, see \cite{brook03}. By observing detailed  chemical abundances of these observed stars, with a program underway at Apache Point and Las Campanas observatories,  we hope to determine whether these high eccentricity, low metallicity, loosely bound stars can be distinguished from field halo stars.

\section{Primordial versus oldest stars: distributions }
\begin{figure}
\begin{center}
\includegraphics[width=4.in]{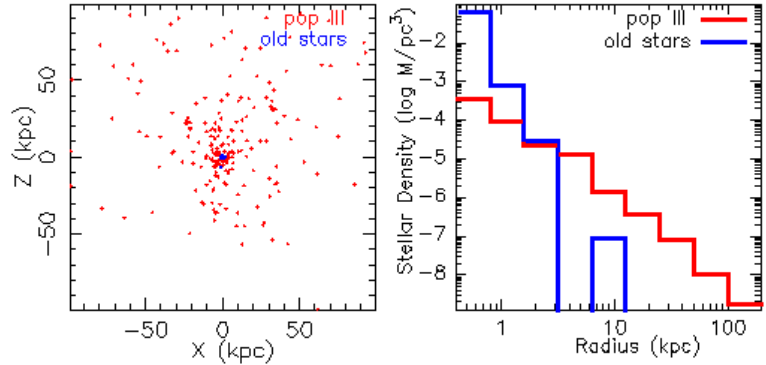}
\caption{Final distribution of the oldest (blue) and metal free (red) stars in the X-Z plane, and their density as a function of radius. Oldest stars form in earliest collapsing, most massive proto-galaxies, and  are located centrally. Metal free stars continue to form in relatively late collapsing, low mass,  isolated,  proto-galaxies and are spread through the halo at $z=0$.}  \label{distribution}
\end{center}
\end{figure}

Where  are the Milky Way's oldest stars to be found? What about primordial stars? If they lived a Hubble time, could we have expected to have observed them? Does our lack of observation of metal free stars constrain their IMF to be top heavy? In figure~\ref{distribution}, we plot the distribution of metal free star particles from our simulation (red) and an equivalent mass of the oldest simulation star particles (blue). 
As expected, the oldest stars, which  form in the proto-galaxies which form in high $\sigma$ peaks which collapse first, are located centrally, primarily in the bulge at $z=0$. Metal free stars, on the other hand, can form in later collapsing, isolated, proto-galaxies in low $\sigma$ peaks. Stars from these low $\sigma$ peaks are spread throughout the stellar halo \cite{brook07}. 

How confident are we that proto-galaxies forming in late collapsing low $\sigma$ peaks will form form from primordial rather than pre-enriched material? In a complimentary study, \cite{scann}, we used analytic feedback prescriptions, coupled with high resolution N-body simulations, to follow the enrichment of the inter galactic medium (IGM), the birth of metal free stars and their final distribution in a Milky Way mass halo. The advantage of this technique is its ability to survey a large range of feedback energies.  Spherical enrichment flows emanating from each collapsed halo (proto-galaxy) are followed at each time-step, thus allowing us to determine whether regions of the IGM have been enriched.    Even with with the largest plausible energy injected, we find that proto-galaxies can form from primordial gas down to $z\sim 5$. 

We conclude that a  different distribution of metal free and oldest stars is expected, driven by combining two results:   difference in density profiles in dark halos from the contribution of different $\sigma$ peaks \citep{moore}, and the fact that the isolated, relatively late forming proto-galaxies (i.e. those forming in low $\sigma$ peaks), form from primordial material. The lack of observed metal free stars in the local region can be taken as constraining the Pop~III IMF to be top heavy, with stars too massive to live a Hubble time. The lowest metallicity observed stars  can be interpreted as holding clues to the chemical yields of Pop~III stars.

\section{Conclusions}
Hierarchical structure formation will leave signatures within the stellar halo of the Milky Way. Different mass proto-galaxies will contribute stars to the halo which will differ in their distributions, and possibly their chemical abundances. A combination of spatial, kinematic and chemical information should allow us to distinguish stars from these different proto-galactic origins. Primordial stars continue to form at relatively late times in  isolated, low mass proto-galaxies. This means that metal free stars will have a more extended distribution than the oldest stars, which form in the earliest collapsing, most massive proto-galaxies, and are centrally located at z=0. Our lack of observations of metal free stars can thus be taken to restrain Pop III star lifetimes to be less than a Hubble time, and the initial mass function of such stars to be top heavy. The lowest metallicity stars observed in the local region contain chemical information pertaining to metal free stars.

\footnotesize

\end{document}